\documentclass[journal=jpclcd]{achemso}%

\usepackage{dcolumn}
\usepackage{bm}
\usepackage[version=4]{mhchem}
\usepackage{siunitx}
\usepackage{subfigure}
\usepackage{amsfonts,amsmath,amssymb, subfigure}
\usepackage{color}
\usepackage{commath}%
\usepackage{esint}%
\usepackage[T1]{fontenc}
\usepackage{graphicx}
\usepackage[utf8]{inputenc}
\usepackage{microtype}
\usepackage{times}
\usepackage[textsize=footnotesize,obeyFinal]{todonotes}
\usepackage{xspace}
\usepackage[version=4]{mhchem}%
\usepackage{textcomp}

\makeatletter
\renewcommand*{\acs@tocentry@print@aux}{%
  \begingroup
    \let\@startsection\acs@startsection@orig
    \acs@section*{\tocentryname}%
    \tocsize
    \sffamily
    \singlespacing
    \begin{center}
          \begin{minipage}{\acs@tocentry@height}
            \vbox to \acs@tocentry@width{\acs@tocentry@text}%
          \end{minipage}%
    \end{center}%
  \endgroup
}
\makeatother
\newcommand*{\origrightarrow}{}

\renewcommand*{\textrightarrow}{\fontfamily{cmr}\selectfont\origrightarrow}

\newcommand{\Qt}{\ensuremath{^3Q_{0+}\,}} 
\newcommand{\Qs}{\ensuremath{^1Q_{1}\,}} 
  
\title{Spectroscopic Signature of Chemical Bond Dissociation Revealed by
Calculated Core-Electron Spectra}
\author{Ludger~Inhester}
\email{ludger.inhester@cfel.de}
\affiliation{Center for Free-Electron Laser Science, DESY, Notkestrasse 85, 22607 Hamburg, Germany}
\altaffiliation{L.I. and Z.L. contributed equally to this work.}
\author{Zheng~Li}
\email{zheng.li@desy.de}
\affiliation{School of Physics, Peking University, Beijing, China}
\alsoaffiliation{Max Planck Institute for the Structure and Dynamics of Matter, D-22761 Hamburg, Germany}
\altaffiliation{L.I. and Z.L. contributed equally to this work.}
\author{Xiaolei~Zhu}
\email{zhuxl@stanford.edu}
\affiliation{Stanford PULSE Institute, SLAC National Accelerator Laboratory, 2575 Sand Hill Road, Menlo Park, CA 94025, USA}
\author{Nikita~Medvedev}
\email{medvedev@ipp.cas.cz}
\affiliation{Institute of Physics Czech Academy of Science  Na Slovance 2, 182 21 Prague 8, Czech Republic}
\alsoaffiliation{Institute of Plasma Physics, Czech Academy of Science, Za Slovankou 4, 182 00 Prague 8, Czech Republic}

\author{Thomas~J.~A.~Wolf}
\email{thomas.wolf@stanford.edu}
\affiliation{Stanford PULSE Institute, SLAC National Accelerator Laboratory, 2575 Sand Hill Road, Menlo Park, CA 94025, USA}

\date{\today}
\begin{document}
\begin{tocentry}
\includegraphics{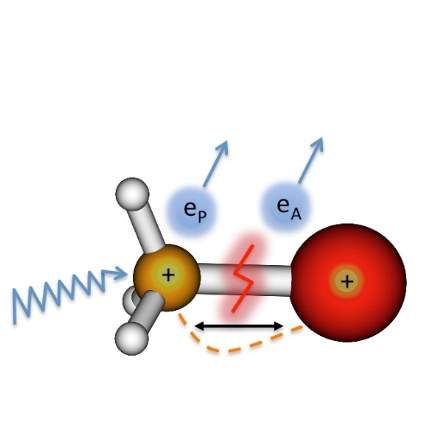}
\end{tocentry}
\begin{abstract}
The advent of ultrashort soft x-ray pulse sources permits the use of established gas phase spectroscopy methods to investigate ultrafast photochemistry in isolated molecules with element and site specificity. In the present study, we simulate excited state wavepacket dynamics of a prototypical process, the ultrafast photodissociation of methyl iodide. Based on the simulation, we calculate time-dependent excited state carbon edge photoelectron and Auger electron spectra. We observe distinct signatures in both types of spectra and show their direct connection to \ce{C-I} bond dissociation and charge rearrangement processes in the molecule. We demonstrate at the \ce{CH3I} molecule that the observed signatures allow us to map the time-dependent dynamics of ultrafast photo-induced bond breaking with unprecedented details.
\end{abstract}


Some of the most important photochemical processes in nature take place in the excited states of organic molecules on ultrafast timescales\cite{cheng_dynamics_2009,polli_conical_2010}. Thorough understanding of ultrafast photochemistry can be achieved by gas phase spectroscopy in combination with high-level ab initio excited state dynamics simulations. Such combined approaches work best, if the experimental and calculated observables can be directly compared as demonstrated for time-resolved valence photoelectron spectroscopy\cite{wolf_hexamethylcyclopentadiene_2014,horton_excited_2019,coates_vacuum_2018,macdonell_excited_2016,schalk_cyclohexadiene_2016,warne_photodissociation_2019}. Time-resolved spectroscopy of the valence electrons, however, is not able to produce site-specific information about photochemistry, e.g. the dissociation of a specific chemical bond. This can be achieved by soft x-ray spectroscopy of the strongly localized inner electrons.
With femtosecond soft x-ray pulses from free-electron laser (FEL) and high-harmonic generation (HHG) sources now becoming available\cite{emma_first_2010,pertot_timeresolved_2017,attar_femtosecond_2017,teichmann_5kev_2016,johnson_highflux_2018,allaria_highly_2012,ackermann_operation_2007}, well established steady-state techniques like x-ray absorption spectroscopy, x-ray photoelectron spectroscopy (XPS), and Auger electron spectroscopy (AES) can be employed to follow ultrafast dynamics of molecules\cite{mcfarland_ultrafast_2014,wolf_probing_2017,wolf_observing_2017,brausse_timeresolved_2018,siefermann_atomicscale_2014,sann_imaging_2016}.

Both steady-state XPS and AES are element specific, since ionization edges of different elements lie hundreds of eV apart. The core electron binding energies, which are measured in XPS, additionally, show a high sensitivity to the local bonding environment of a specific atom in a molecule\cite{siegbahn_esca_1969,bagus_mechanisms_1999}. 
Thus, time-resolved XPS can be expected to be sensitive to changes in the bonding environment induced by ultrafast photochemistry. Photoionization of a $1s$ electron as in XPS leaves the molecule with a strongly localized electron hole that typically decays within few femtoseconds via Auger decay creating two holes in the valence.  
Because of the Coulomb repulsion of the two valence holes, also time-resolved AES can be expected to be sensitive to structural changes in the molecule \cite{inhester_auger_2012,mcfarland_ultrafast_2014,marchenko_ultrafast_2018}. 
The interplay of molecular effects accompanying core ionization give rise to valence rearrangement between the different atoms that are of strong fundamental interest\cite{erk_imaging_2014,boll_charge_2016,rudenko_femtosecond_2017,hao_theoretical_2019}. 
Investigating the transition from an intact molecule to a dissociated fragment via femtosecond-resolved XPS and AES holds the potential for elucidating important aspects of these mechanisms.

%
Here, we present a theoretical investigation of the TR-XPS and TR-AES signatures of a prototypical process, the ultrafast photodissociation of the carbon-iodine bond in isolated methyl iodide. This reaction has been experimentally investigated in several studies\cite{denalda_detailed_2008,corrales_control_2014,erk_imaging_2014,boll_charge_2016,brausse_timeresolved_2018,tehlar_timeresolved_2013,warne_photodissociation_2019}. The molecule is therefore an ideal benchmark for the capabilities of time-resolved XPS and AES. Photoexcitation of methyl iodide in the ultraviolet (UV) spectral regime leads to a single electron transition from an iodine-centered $\pi$ molecular orbital (MO) to a carbon-iodine antibonding MO ($\pi \to \sigma^*$ excitation). The dipole-allowed spin-orbit coupled \Qt and \Qs states contribute dominantly to the UV photoabsorption. For the \SIrange{260}{262}{\nm} region (absorption maximum for \Qt state), the triplet \Qt state accounts for 94\% of the absorption intensity\cite{gedanken_magnetic_1975}. Both states are repulsive and \ce{C-I} dissociation is known to take place on a sub-\SI{100}{\fs} timescale\cite{denalda_detailed_2008,corrales_control_2014,hammerich_timedependent_1994,erk_imaging_2014,attar_direct_2015,drescher_communication_2016,boll_charge_2016,baumann_timeresolved_2018}. 
The photodissociation dynamics of methyl iodide have been experimentally investigated by XPS in a seminal study at the iodine $4d$ edge \cite{brausse_timeresolved_2018}. The study identifies a time-dependent shift of the iodine $4d$ binding energy as the XPS signature of C-I bond dissociation. With our present study, we theoretically investigate the XPS and AES during photochemical C-I bond breaking in \ce{CH3I} using ab initio wavepacket simulations combined with a complete description of the electronic structure in the photoionization and subsequent Auger decay process employing the XMOLECULE toolkit\cite{hao_efficient_2015,inhester_xray_2016,inhester_chemical_2018}. We demonstrate that the transient XPS signatures of bond dissociation can be considerably richer than the so far experimentally observed linear shift in binding energy. As opposed to the aforementioned experimental study, we are simulating carbon K-edge XPS and AES signatures. Other than the iodine $4d$ edge, the carbon K-edge features do not exhibit extensive fine structure from spin-orbit coupling. This facilitates the interpretation of time-dependent spectroscopic signatures. We show that TR-XPS is highly sensitive to the immediate electron density redistribution during bond dissociation dynamics and electron rearrangement upon core-shell ionization. The sensitivity to electron density redistribution is enabled by dramatic changes in the ionization transition strength to different core-ionized states during the photodissociation, which has been to the best of our knowledge never been considered before.
Thus, both time-resolved XPS and AES exhibit a distinct signature, when the redistribution of bonding electron density is completed.
%
These signals can be utilized as a probe for molecular valence electron rearrangement accompanying core-shell ionization.

\begin{figure}
\subfigure[Snapshots of the nuclear wave packet density]{\includegraphics[width=0.5\textwidth]{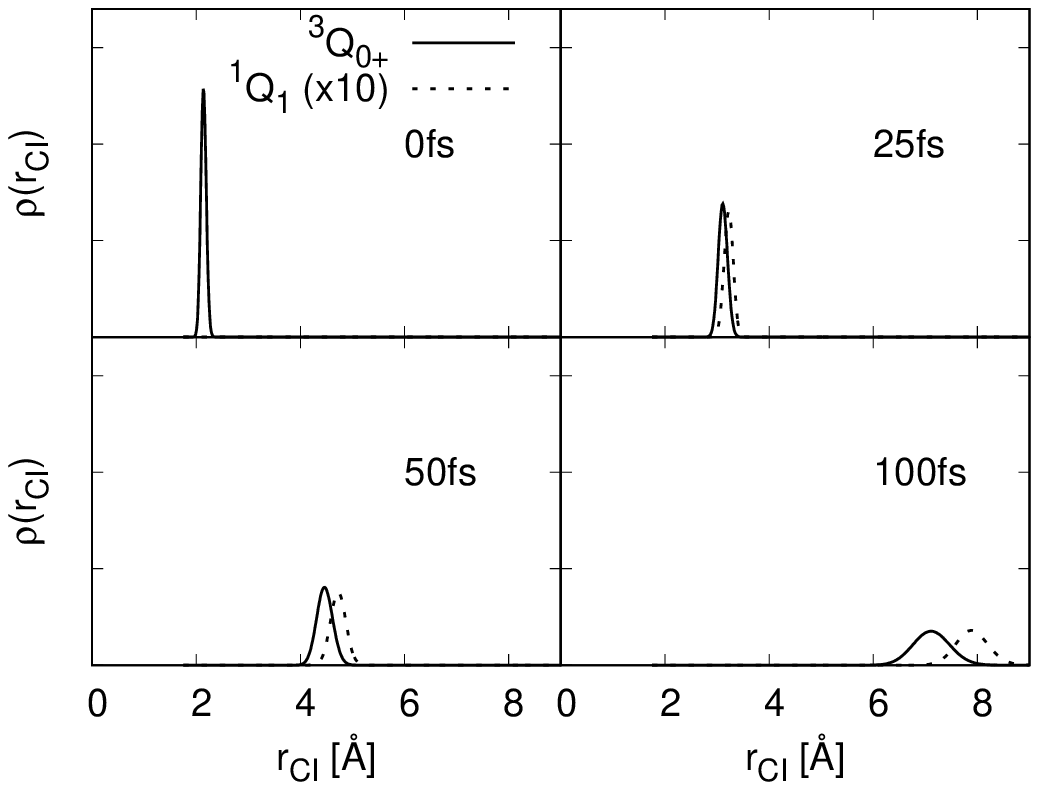}}%
\subfigure[Electronic state-population]{\includegraphics[width=0.5\textwidth]{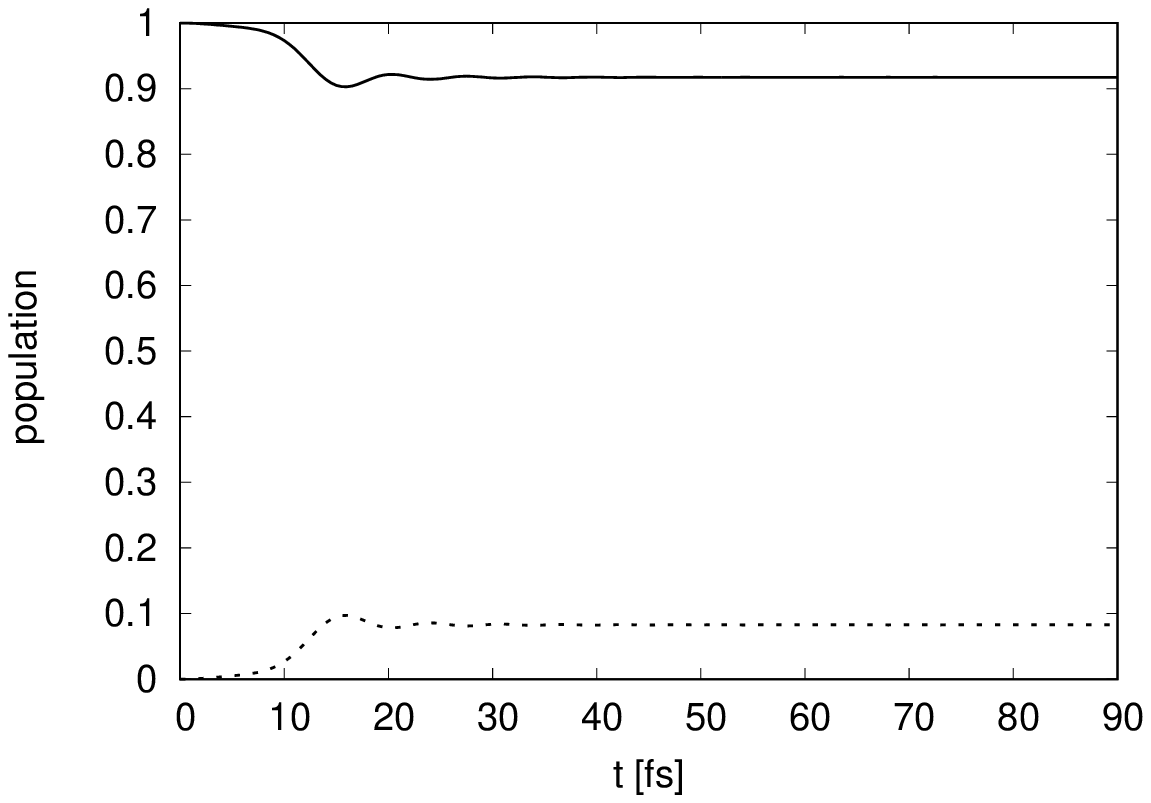}}
\caption{Wave packet dynamics of \ce{CH3I}: (a) The reduced density  of the nuclear wave packet on the \Qt state (solid line) and the \Qs state (dotted line, magnified by factor 10).
(b) Evolution of the population in the \Qt and \Qs state. 
\label{fig:quantum_dynamics}}
\end{figure}
The evolution of the molecule in the transient excited state is shown in Fig.~\ref{fig:quantum_dynamics}(a-b).
Snapshots of the reduced density of the wave packet along the \ce{C-I} bond distance that is obtained by projecting the wave packet onto one of the electronic states and tracing out all other nuclear degrees of freedom are shown in Fig.~\ref{fig:quantum_dynamics}(a). 
Within $\simeq\SI{50}{\fs}$, the bond distance is increased to two times the ground state equilibrium value. 
Due to the strongly repulsive potential, the wave packet stays relatively compact throughout the first \SI{100}{\fs} after UV excitation. 
Additionally, the \ce{C-H_3} umbrella motion (not shown) brings the \ce{CH3} fragment from the tetrahedron structure to a closely planar geometry as the \ce{C-I} bond dissociates.
Some of the electronic state population is transferred to the lower \Qs state through non-adiabatic effects, 
but the molecule remains almost completely in the \Qt state\cite{hammerich_timedependent_1994} 
(see Fig.~\ref{fig:quantum_dynamics}(b)).
For all considered bond lengths, the spin-orbit coupled \Qt state is dominated by the $1^3E$ spin configuration.
While for the considered UV wavelength spin-orbit effects are relevant for the photoexcitation\cite{gedanken_magnetic_1975} and for the iodine state into which the molecule dissociates, they play a minor role for the electronic state of the corresponding methyl fragment\cite{denalda_detailed_2008,baumann_timeresolved_2018,kamiya_initio_2019, corrales_control_2014}. For the XPS and AES due to carbon 1s ionization, spin-orbit splitting effects  are smaller than $\SI{1}{\eV}$\cite{cutler_ligand_1992} and thus play a minor role. Accordingly, we neglect in the following discussion spin orbit-coupling and discuss the time dependent spectra for the non-spin-orbit coupled $1^3E$ state.

\begin{figure}
\subfigure[XPS as a function of bond distance]{\includegraphics[width=0.70\textwidth]{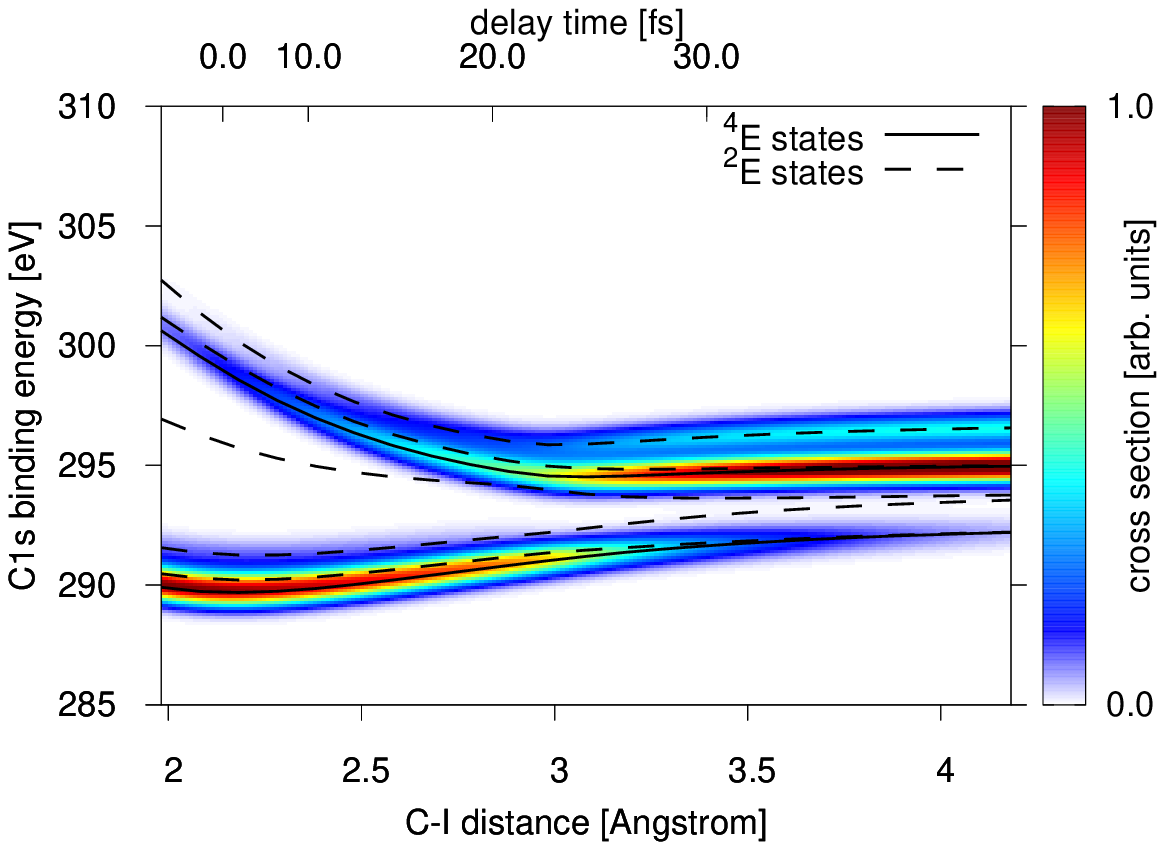}}\\
\subfigure[XPS as a function of delay time]{\includegraphics[width=0.70\textwidth]{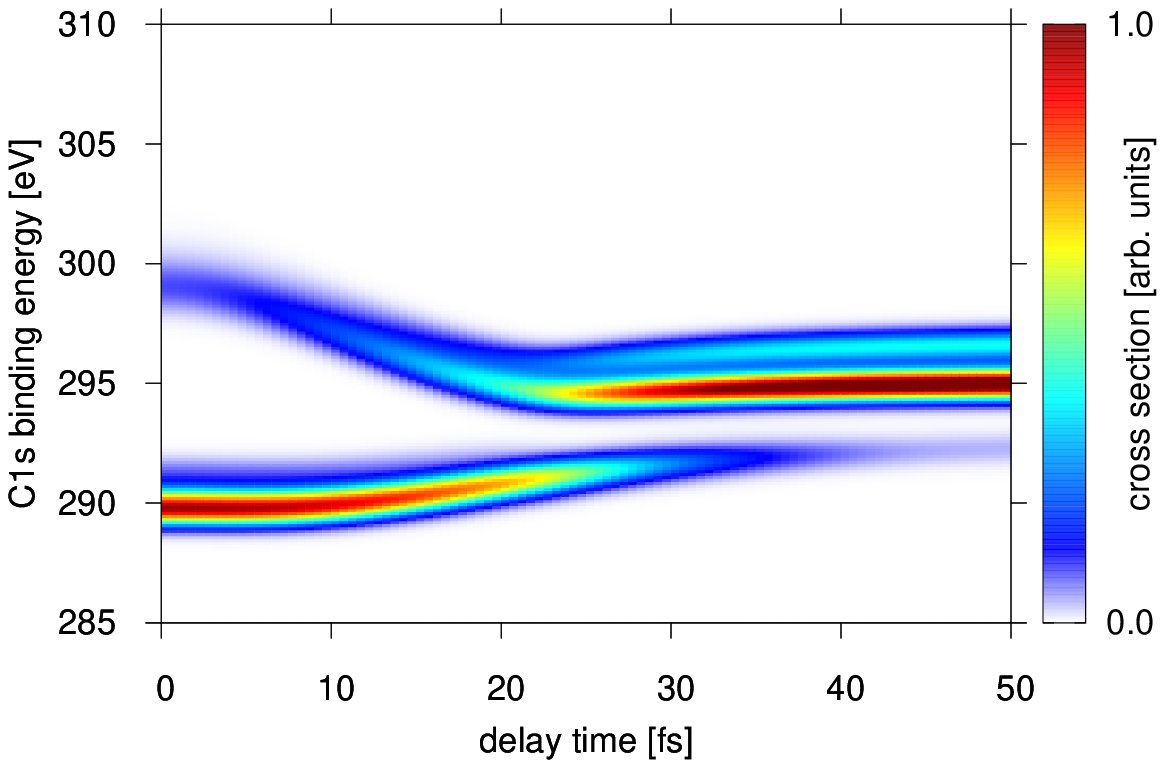}}
\caption{Femtosecond XPS of the UV-excited \ce{CH3I} molecule ($1 ^3E$ state)
as a function of \ce{C-I} distance (a) and as a function of  delay time to the initial pump step (b). 
The hydrogen positions have been fixed at their equilibrium position.
The incident photon energy is assumed to be \SI{350}{\eV}, and a spectral width of \SI{1}{\eV} (full-width at half-maximum) is assumed to convolve the calculated XPS transitions. 
In the upper plot, solid and dashed lines mark the difference of the initial $^3E$ potential energy curve to $^4 E$ and $^2 E$ core ionized potential energy curves, respectively. The upper axis mark the center of the wave packet at selected delay times.
\label{fig:xps}
\label{fig:xps_time}
}
\end{figure}
In Fig.~\ref{fig:xps}(a), we present the calculated XPS for the photo-excited \ce{CH3I} molecule as a function of bond-length assuming a photon energy of \SI{350}{\eV} convoluted with a fixed-width Gaussian of \SI{1}{\eV} full-width-at-half-maximum (FWHM). 
At the equilibrium geometry of the ground state, the C$1s$ binding energy is $\simeq \SI{292}{\eV}$ for the singlet ground state. After excitation into the $1^3E$ state, the C$1s$ binding energy lowers to $\simeq \SI{290}{\eV}$. 
The center of the XPS band observed at $\simeq \SI{290}{\eV}$ increases slightly 
with bond length up to an internuclear distance of \SI{3}{\angstrom}.
A satellite feature that can be attributed to valence shake-up is seen very weakly at equilibrium bond distance (\SI{2.2}{\angstrom}) at photoelectron energy $\simeq \SI{300}{\eV}$. For larger bond distances this satellite feature becomes stronger and its binding energy decreases.
For interatomic separations larger than 
$\simeq \SI{3}{\angstrom}$, the initial weak satellite feature, which is at this bond distance at $\simeq \SI{295}{\eV}$,
becomes dominant and the original photoelectron line vanishes. While for the equilibrium bond distances very small splitting between doublet and quadruplet final electronic states is seen, at bond distances $> \SI{3.5}{\angstrom}$, distinct quadruplet and doublet features with an energy separation of $\simeq \SI{1.9}{\eV}$ occur. 
Due to their larger multiplicity, the two final quadruplet states (solid lines in Fig.~\ref{fig:xps}(a)) dominante the TR-XPS with a relative strength twice as large as the corresponding doublet states.
The TR-XPS spectrum resulting from the time-dependent wave packet (Fig.~\ref{fig:quantum_dynamics})
is shown in Fig.~\ref{fig:xps_time}(b). In this time-resolved spectrum, we have neglected the effect of the relatively small hydrogen motion that has little impact on the resulting x-ray spectra.
The similarity of time and distance-dependent XPS showcases direct mapping onto the dissociation reaction coordinate.
As can be seen, at a time delay of $\simeq \SI{20}{\fs}$ the position of the dominant photoelectron spectral line shows a pronounced jump that is connected with the wave packet being located at bond distances around \SI{3}{\angstrom}.

\begin{figure}
\includegraphics[width=0.8\textwidth]{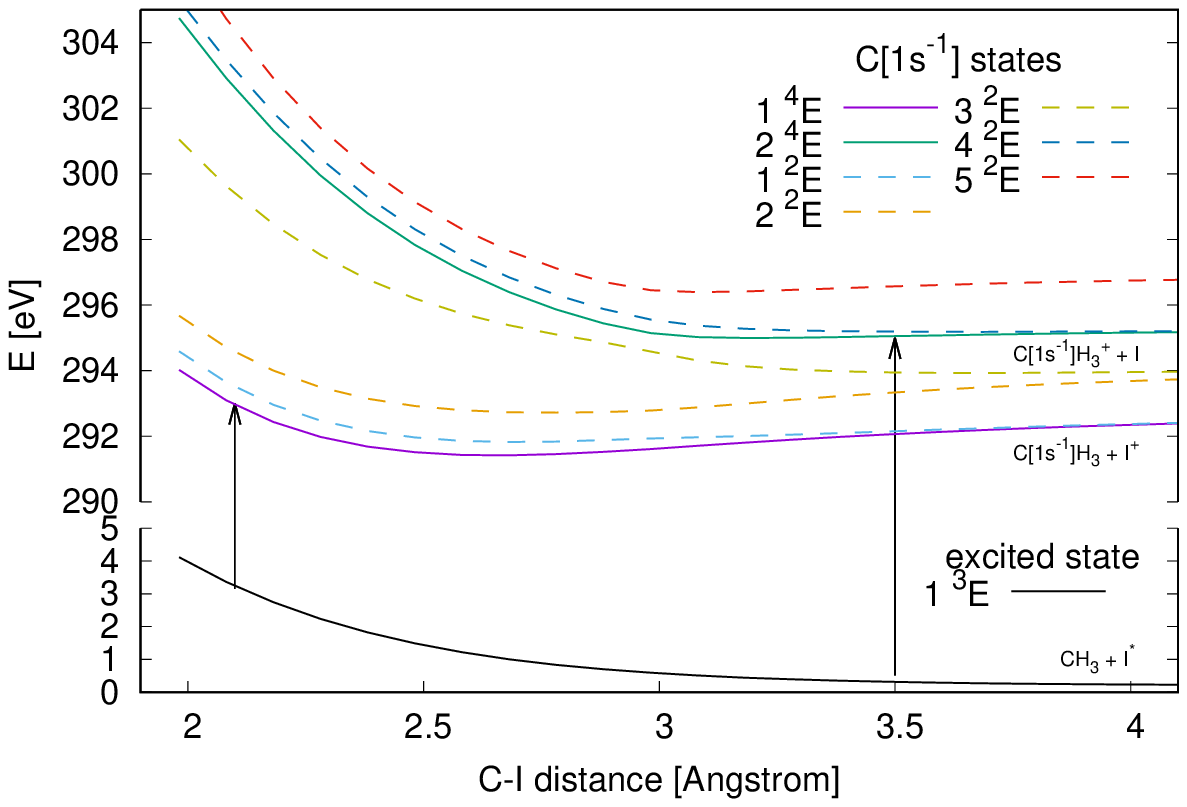}
\caption{Potential energy curves of UV-excited \ce{CH3I} molecule ($1^3E$ state) 
and carbon-core-ionized states with different valence configuration 
(only states of irreducible representation E with doublet and quadruplet symmetry
are shown).
\label{fig:pec}
}
\end{figure}
The spectral position of the photoelectron lines follow the difference of the
potential energy curves of the excited and core ionized potential energy curves shown in Fig.~\ref{fig:pec}.
The jump in the XPS at a bond length~$\simeq \SI{3}{\angstrom}$ is a consequence of the bond-length dependency of the core ionization cross-section to the different core-hole states.
\begin{figure}
\includegraphics[width=0.8\textwidth]{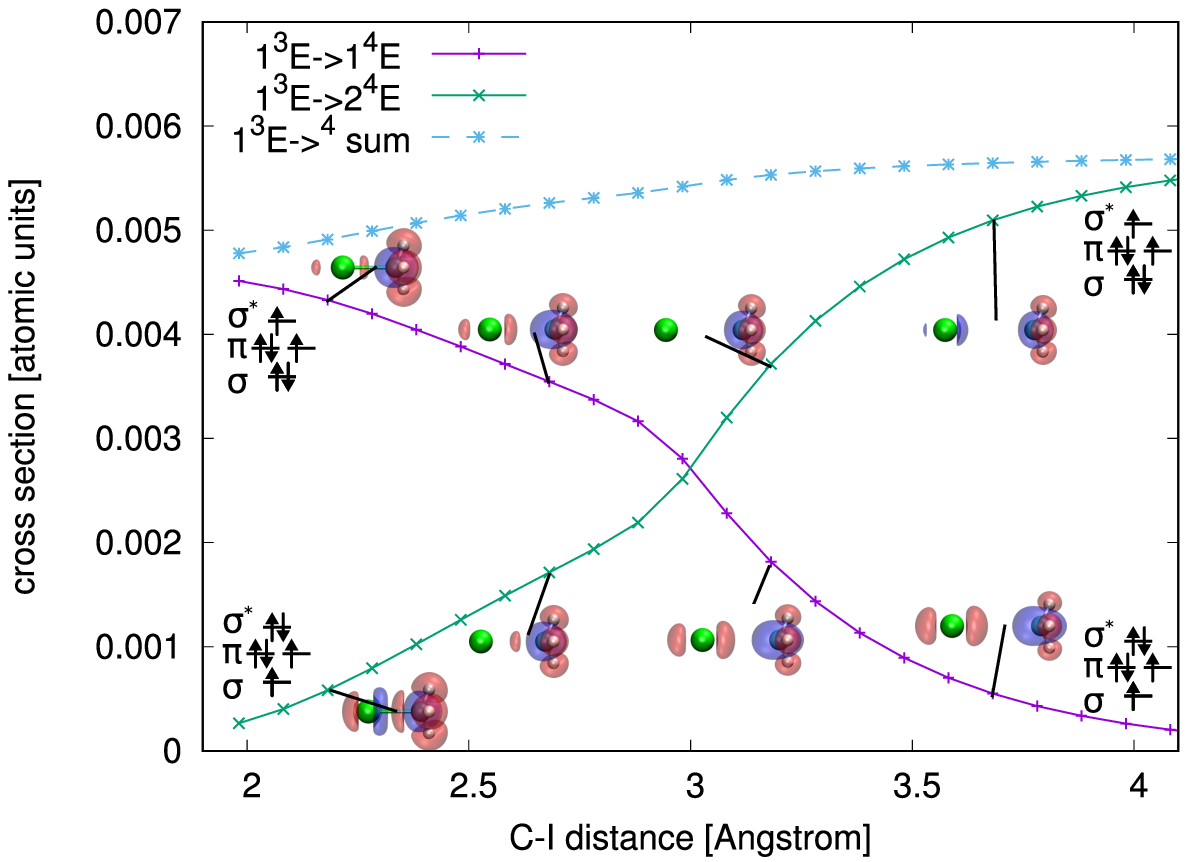}
\caption{Cross-sections for photoionization from the excited state $1^3E$ to the carbon core-ionized $1^4E$ and $2^4E$ states
as a function of inter-atomic separation.
The upper axis indicates the center of the nuclear wave packet for selected delay times.
The blue dashed line shows the sum of both transitions.
For selected distances, iso-surface plots show the electron density change associated with the core ionization.
The blue color indicates electron density increase, the red color indicates electron density decrease upon core ionization.
For the limiting cases (not dissociated/ fully dissociated), we depict the corresponding valence orbital occupation scheme.
\label{fig:cs}
}
\end{figure}
This is clearly illustrated in Fig.~\ref{fig:cs}, which shows the swap in the 
individual photoionization cross sections for the two involved final core-hole states of quadruplet spin multiplicity. 
The lower core-ionized state $1^4E$, which is dominantly populated through core shell ionization at equilibrium bond length, 
describes a redistribution of the valence electron density that screens the created core hole.
It is instructive to realize that this lower carbon-core ionized state, which can be seen as a satellite for photoionization at large interatomic distance, corresponds to an asymptotic dissociation into 
I$^+$ and a core-excited \ce{CH3} radical.
The same counter-intuitive dissociation character of core hole states, where the charge is located on the atom that has not been core-ionized,
has also been reported for CO\cite{corral_potential_2017}.
On the contrary, the upper core-ionized quadruplet state $2^4E$ corresponds to an asymptotic dissociation into neutral \ce{I} and a core-ionized \ce{CH3} cation.
The electron density change accompanying core ionization of the initial photoexcited state is illustrated
in the iso-surface plots in Fig.~\ref{fig:cs} for selected \ce{C-I} distances.
The blue iso-surface (electron density increase) around the carbon atom and the red iso-surfaces (electron density decrease) 
covering the hydrogen atoms clearly reflect the contraction of the valence electron density in the methyl moiety 
illustrating the screening of the core-hole.
Additionally, one can see a shift of valence electron density involving the iodine atom that is qualitatively different for the 
two considered core-ionized states.
The $1^3E \to 1^4E$ transition tends to shift electron density from the iodine towards the methyl moiety.
In contrast, density change due to the $1^3E \to 2^4E$ transition is for small bond lengths 
characterized by a reshuffeling of $\sigma$ density. For larger bond lengths, it shows almost no impact on the 
iodine atom.
We note that the here discussed core-ionized or core-excited \ce{CH3} fragments are in transient states that
will further charge up via Auger decay.

The characteristic valence electron changes following core ionization can be directly understood from 
the configurational mixing in the valence electronic structure as a function of interatomic separation.
While the photoexcited $1^3E$ state is for all bond distances characterized by a $\pi \to \sigma^*$ excitation,
the two core hole states $1^4E$ and $2^4E$ show a mixing of core-hole configurations involving an 
additional $\sigma \to \sigma^*$ excitation that reflects the tendency in the core ionized states to
screen the core hole via shifting valence electrons from the iodine atom towards the methyl moiety.
At equilibrium bond length, the $1^4E$ state is dominated by the core-ionized $\pi \to \sigma^*$ configuration.
The core hole leads to a somewhat stronger bond and, thus, to a slightly less dissociative potential energy curve compared to 
the initial valence excited state $1^3E$. 
Accordingly, the core-electron binding energy slightly increases with \ce{C-I} distance (see Fig.~\ref{fig:xps}(a)).
This observation of strengthening of a bond through a core hole has been observed for a number of other core ionized molecules\cite{corral_potential_2017}.
With increasing bond length, the configurational mixing of the $1^4E$ core-hole state is shifted more and more towards an additional $\sigma \to\ \sigma^*$ excitation,
which becomes dominant for bond length $> \SI{3}{\angstrom}$. Due to this change in configurational character the photoionization from the initial $\pi \to \sigma^*$ excited state is depleted.
The higher-lying $2^4E$ core-ionized state is for small bond distances dominated by the additional $\sigma \to \sigma^*$ excitation resulting in a potential energy curve that is even more repulsive than the initial $1^3E$ excited state (see Fig.~\ref{fig:pec}).
This is also indicated by the lowering of the respective photoelectron binding energy with 
inter-atomic distance (see Fig.~\ref{fig:xps}(a)).
The configurational character of this state changes with bond length in the opposite way: 
At the Franck-Condon region it describes a $\sigma \to \sigma^*$ shake-up and for bond length $>\SI{3}{\angstrom}$
it is characterized by the main $\textrm{C}(1s^{-1}),\pi \to \sigma^*$ configuration (i.e. without additional $\sigma\to\sigma^*$ excitation).
Accordingly, photoionization from the initial $\pi \to \sigma^*$ excited state increases with bond length.

From this finding we can directly connect the jump in the XPS with the suppression of the inter-atomic 
valence electron rearrangement in the context of core-hole screening.
These dynamics can also be understood with the classical over-the-barrier model of charge migration\cite{erk_imaging_2014,ryufuku_oscillatory_1980} (see the supplementary material for details). 
According to this model, the electron transfer from \ce{I} to \ce{CH_3^+$[1s^{-1}]$} is forbidden, 
when the internuclear distance is larger than the critical distance $R_c=\SI{3.3}{\angstrom}$, which is consistent
with the observed critical distance of $R_c\simeq\SI{3.0}{\angstrom}$ in the XPS (Fig.~\ref{fig:xps_time}(a)).

\begin{figure}
\subfigure[AES as a function of bond distance]{\includegraphics[width=0.8\textwidth]{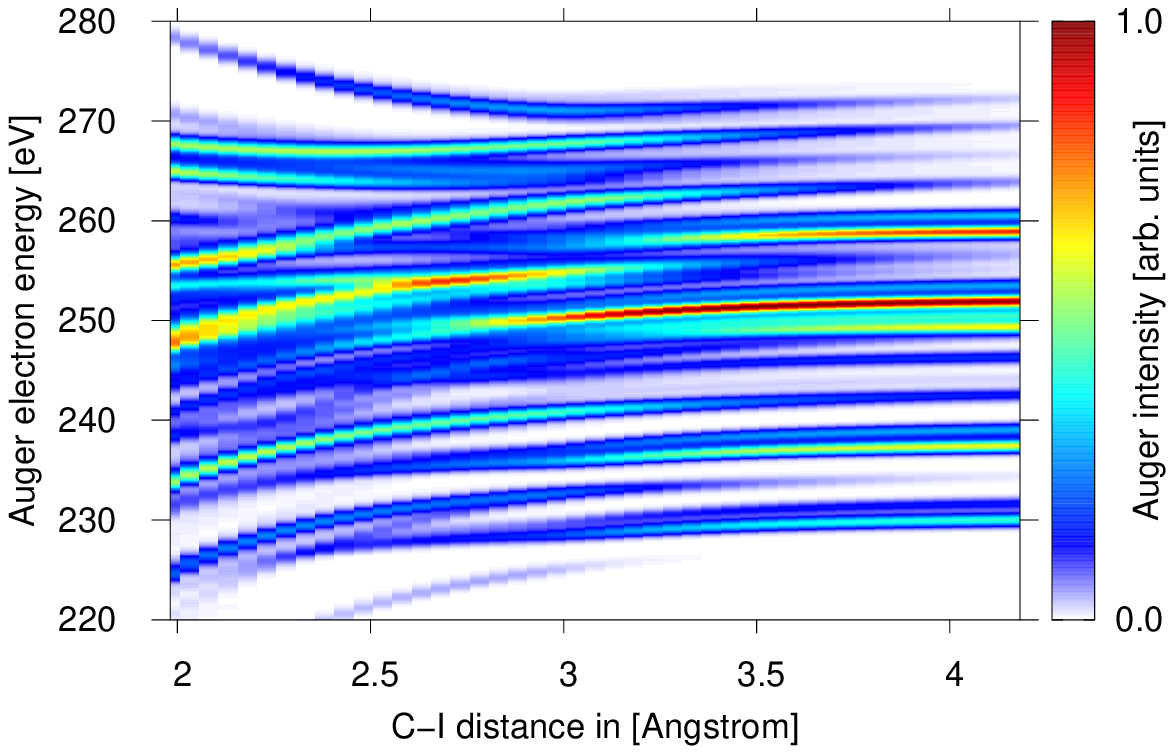}}
\subfigure[AES (high energy region) for selected bond lengths]{\includegraphics[width=0.70\textwidth]{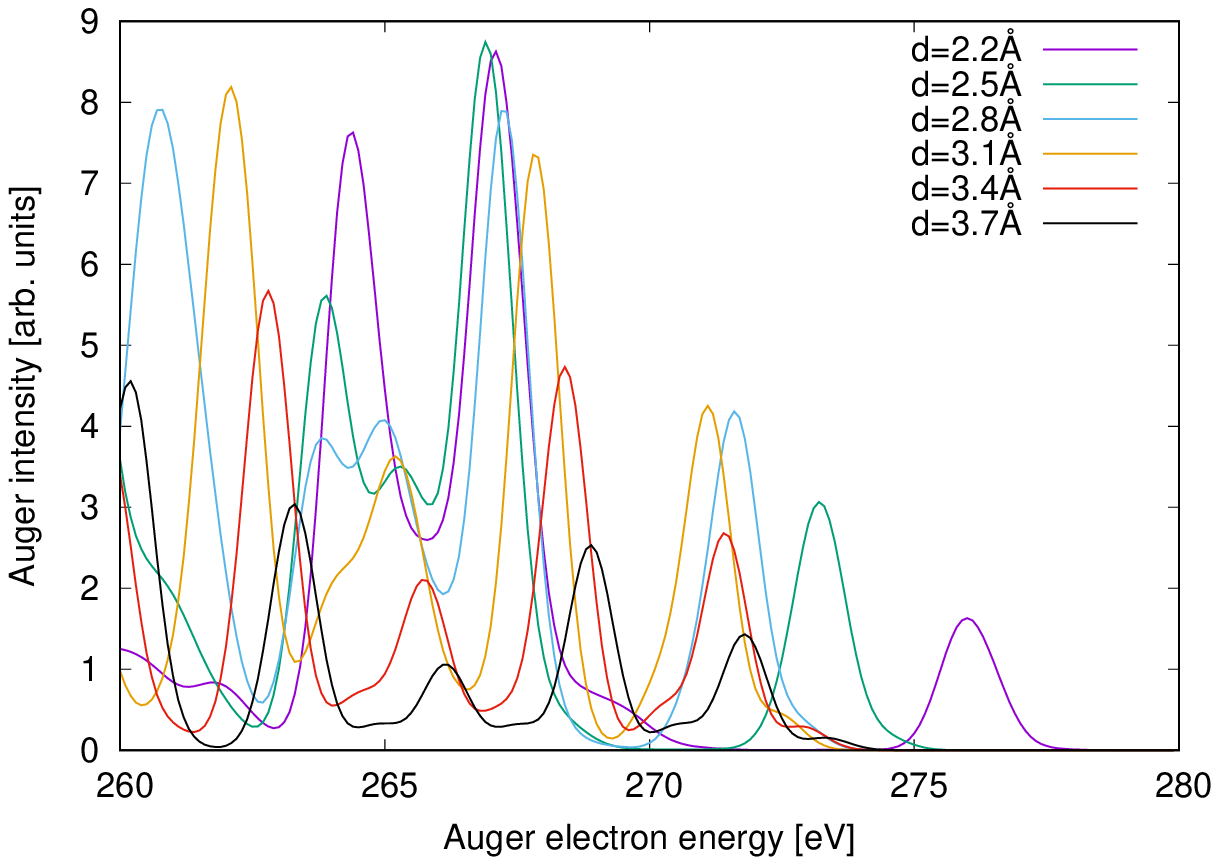}}
\caption{Femtosecond AES of UV excited and carbon-core ionized \ce{CH3I} molecule as a function of photoelectron binding energy and \ce{C-I} bond length. 
The hydrogen positions have been fixed at their equilibrium values.
The incident photon energy is assumed to be \SI{350}{\eV}, and a spectral width of \SI{1}{\eV} (full-width at half-maximum) is assumed to convolve the calculated Auger transitions. 
Figure 5a shows the full scan as a function of bond length and Auger electron energy.
Figure 5b shows the high energy part of the Auger spectrum for selected bond lengths.
\label{fig:aes}
}
\end{figure}

Strong sensitivity with respect to bond dissociation is also seen in the TR-AES that is shown in Fig.~\ref{fig:aes}(a+b). 
Because of the multitude of different pathways resulting in densely lying Auger lines,
a complete interpretation of all the lines and their bond distance dependence is tedious.
There is a general trend for most of the Auger lines to show a shift to higher energy with increased inter-atomic separation.
This effect can be attributed to the possible larger spatial separation and thus smaller Coulomb repulsion of the two valence holes in the final dicationic states that dissociate into two ion pairs.
Around \SI{3}{\angstrom}, many lines show a discontinuity that is a consequence of the discussed swap in the photoionization cross sections (Fig.~\ref{fig:cs}).
Above \SI{3.5}{\angstrom}, the Auger lines seem to vary very weakly with interatomic distance.
At these bond-lengths, the Auger decay in the \ce{CH3} fragment seems to be almost unaffected by the iodine fragment.

In the following we focus on the Auger line resulting in the fastest Auger electrons.
Typically, these fast Auger electrons are rather isolated from other Auger lines, because 
they represent low energetic dicationic states, where the density of states of the dication is rather low compared to  higher excited dicationic states.
In general, fast Auger electrons are therefore easier to identify with specific dicationic electronic states.
As shown in Fig.~\ref{fig:aes}(a+b), the Auger spectrum exhibits a satellite feature above \SI{270}{eV} that becomes stronger 
around a \ce{C-I} distance of \SI{3}{\angstrom} and vanishes again for larger bond lengths.
At equilibrium bond length $\simeq \SI{2.2}{\angstrom}$, this Auger line is found at energies $\simeq\SI{277}{eV}$ 
and its energetic position lowers with increasing \ce{C-I} distance.
Our calculation indicates that this Auger transition can be attributed to the decay of the higher-lying core-ionized quadruplet state $2^4E$ into the dicationic $1^3E$ state. The change of energetic position of this Auger line is dominated by the strongly repulsive curve of the core-ionized $2^4E$ state.
The dicationic $1^3E$ state associated with this Auger channel 
dissociates into \ce{I^+} and \ce{CH3^+}.
Thus, this Auger channel is accompanied with a transfer of a valence electron to the iodine atom. Analogous to the blocking of charge transfer in the XPS the intensity of this Auger line becomes therefore weaker with increasing bond length. 

%

Via ab-initio modelling of XPS and AES, we have demonstrated here the power of element and site-specific soft x-ray spectroscopy for investigation of photoinduced bond dissociation, a fundamental process in photochemistry. We show at the example of methyl iodide that carbon $1s$ time-resolved x-ray photoelectron spectra exhibit a distinct jump in between the initial core ionization potential of the \ce{CH3I} molecule and of the reaction product \ce{CH3}. As we discuss in the Supp. Information, the jump is clearly preserved when finite time resolution of the experiment is taken into account.
We connect this jump to the point in time and space (\SI{3.0}{\angstrom} and 20 fs), at which molecular charge redistribution upon carbon core ionization is blocked. The large sensitivity of XPS can be understood from the fact that 
the creation of a core hole is accompanied with screening effects in the valence shell, which
lead to rearrangement of valence electrons from the electron-rich iodine atom to the \ce{CH_3} moiety. The reported behavior also reveals that the production of $\mathrm{I}^+$ fragments should show a strong sensitivity with delay time resulting from the different degree to populate the two different states through carbon core ionization. Thereby, our calculation present an interesting explanation of the bond-distance dependence of the molecular charge rearrangement 
mechanism upon x-ray ionization\cite{erk_imaging_2014,boll_charge_2016}.
In comparison to our simulated XPS, the information content from time-resolved valence photoelectron spectroscopy is fundamentally different. The only such study on \ce{CH3I} in the literature employs insufficient probe photon energies to follow the bond dissociation to the photoproducts. \cite{warne_photodissociation_2019}. With sufficient probe photon energy, we can, however, expect to observe features, which correlate with the lowest cationic states of the photofragments \ce{CH3} and \ce{I^*}, to smoothly evolve on the dissociation coordinate from their binding energies in \ce{CH3I} towards their binding energies in the fragments\cite{berkowitz_photoionization_1981,berkowitz_three_1994}. Instead, our simulated XPS provides exclusive information on the evolution of the valence electronic structure of the emerging \ce{CH3} fragment. Moreover, the valence photoelectron spectrum can not be expected to exhibit such a drastic change in the partial ionization cross-sections as we observe in the XPS. 
The TR-AES shows a remarkable sensitivity to \ce{C-I} bond length. Although the spectrum is overall very crowded, particular features shift by more than $\SI{5}{eV}$, which should result in a clearly observable signal in a time-resolved AES experiment.
It is remarkable that in such a pump-probe experiment, specific transient features are only visible in a region of intermediate \ce{C-I} bond distance. Within a UV-pump-x-ray-probe scheme, this feature could be used 
as a sensitive indicator for a specific dissociation step.

While the predicted phenomena have not yet been fully resolved experimentally, \cite{brausse_timeresolved_2018} a confirmation of our predictions is well within the range of capabilities at current free electron laser facilities. We, furthermore, do not see any reason why they should be confined to our specific benchmark molecule and expect them to be general signatures of valence electron rearrangement during bond dissociation.
With the help of the here reported spectral fingerprints of bond-breaking
and new achievements in providing short and accurately controlled x-ray pulses,
we expect time-resolved x-ray photoelectron spectroscopy and time-resolved Auger electron spectroscopy to become an enormously
useful tool to investigate ultrafast photochemistry.

\section{Methods}
\subsection{Quantum dynamics of \ce{CH3I} photodissociation}
In order to describe the non-Born-Oppenheimer nuclear dynamics triggered by photoexcitation of methyl iodide, we performed a five-dimensional quantum dynamics simulation using the multi-configuration time-dependent Hartree (MCTDH) method\cite{beck_multiconfiguration_2000} reproducing the results of Ref.~\citenum{hammerich_timedependent_1994}.
In this work, the numerical procedure demonstrated to faithfully reproduce the branching ratio of I/I* and quantum states distribution of dissociation products. Since it is well described elsewhere\cite{beck_multiconfiguration_2000,hammerich_timedependent_1994}, we here only give a brief overview.

\begin{figure}
\includegraphics[width=0.5\textwidth]{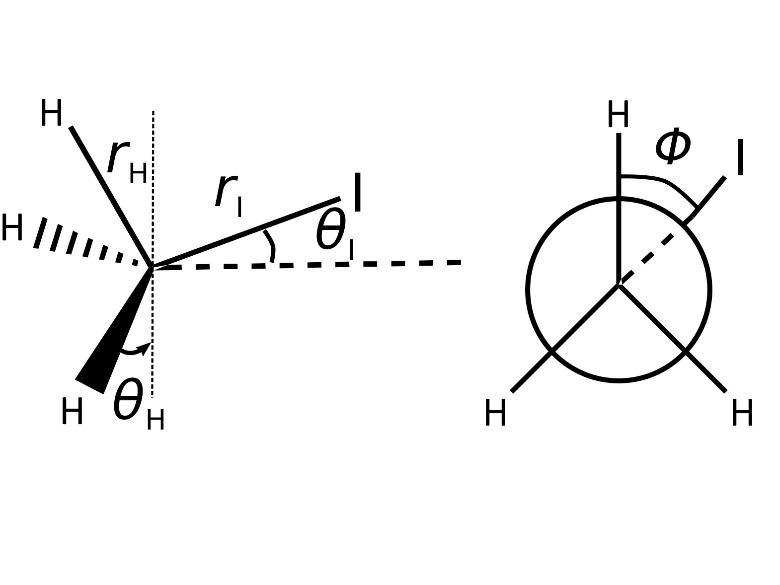}
\caption{\label{fig:coordinates}%
  Coordinates set $R=\{r_I,r_H,\theta_I,\theta_H,\phi\}$ for the five-dimensional quantum dynamics.}
\end{figure}

A set of coordinates is used to describe the processes depicted in Fig.~\ref{fig:coordinates}, the \ce{C-I} dissociation, \ce{C-H} vibration, \ce{H_3-C-I} bending, umbrella motion of \ce{CH3} and wagging of \ce{I} atom. 
In the chosen coordinate space, $r_{I}$ is the major coordinate characterizing the dissociation process.
Motion of $\theta_{{H}}$ and $r_\text{H}$ are driven naturally by the recoil of C-I dissociation.
$\theta_{{I}}$ is crucial to break the $C_{\text{3v}}$ symmetry of the molecule to enable the non-Born-Oppenheimer coupling between the valence excited states, which leads to the formation of excited and ground state iodine atoms, and $\phi$ plays minor role. The initial state of the wave packet of the UV excited \ce{CH3I} molecule is prepared by sudden excitation approximation following the Franck-Condon principle~\cite{hammerich_timedependent_1994}.
The employed potential energy surfaces for the \Qt and \Qs state are based on ab initio spin-orbit configuration interaction calculations~\cite{amatatsu_initio_1991,hammerich_timedependent_1994}.

The time-dependent wave function of the \ce{CH3I} molecule is represented as
\begin{eqnarray}
  &&\chi\left(Q_{1}, \cdots, Q_{p}, t\right)=\sum_{j_{1}=1}^{N_{1}} \cdots \sum_{j_{p}=1}^{N_{p}} A_{j_{1} \cdots j_{p}}(t) \prod_{\kappa=1}^{p=6} \varphi_{j_{\kappa}}^{(\kappa)}\left(Q_{\kappa}, t\right)\nonumber\\
  &=&\sum_{J} A_{J} \Phi_{J}
  \,,
\end{eqnarray}
for our case $N_p=N_6=2$, where $R=\{r_I,r_H,\theta_I,\theta_H,\phi\}=(Q_1,\cdots,Q_5)$ and $Q_6 $ denotes the discrete basis of the electronic degrees of freedom.
The function $\varphi_{j_{x}}^{(\kappa)}\left(Q_{K}, t\right)$ are the time-dependent single particle function (SPF) basis that forms a co-moving cover of the wave packet. 
The equation of motion used to propagate the SPF and the expansion coefficients $A_{J}$ in time follow from the Dirac-Frenkel variation principle as
\begin{equation}
  i \dot{A}_{J}=\sum_{L}\left\langle\Phi_{J}|H| \Phi_{L}\right\rangle A_{L}
  \,,
\end{equation}
and
\begin{equation}
i \dot{\varphi}^{(\kappa)}=\left(1-P^{(\kappa)}\right)\left(\rho^{(\kappa)}\right)^{-1}\langle H\rangle^{(\kappa)} \varphi^{(\kappa)}
\,,
\end{equation}
where the vector notation
$\varphi^{(\kappa)}=\left(\varphi_{1}^{(\kappa)}, \cdots, \varphi_{N_{k}}^{(\kappa)}\right)^{T}$
has been used. 
Further details on the MCTDH method and the the definitions of the
mean field $\langle H\rangle^{(\kappa)}$, the density matrix $\rho^{(\kappa)}$ and the projector $P^{(\kappa)}$ can be found in Ref.~\citenum{beck_multiconfiguration_2000}.

\subsection{Femtosecond XPS and AES of the photodissociating molecule}
For simplicity, we discuss the resulting XPS and AES without the effect of spin-orbit interaction.
As discussed in the result section, the most dominantly populated \Qt state can be associated almost purely with a non-spin-orbit coupled $1^3 E$ state.
The resulting XPS might be blurred due to splitting of the final, core-ionized states.
We estimate the effect of splitting due to spin-orbit splitting to be smaller than $\SI{1}{eV}$\cite{cutler_ligand_1992}.
Given the additional spreading of the spectra due to the spatial extend of the vibrational wave-packet on the dissociating potential energy curves,
spin-orbit splitting details will most-likely be hidden within a broad peak of the individual photo- and Auger electrons.
The spin-orbit coupling therefore does not involve any characteristic changes in the XPS and AES and for our discussion it is therefore sufficient to neglect spin-orbit interaction.

For a given molecular geometry, the photoionization cross section of the initial electronic state $I$ can be calculated as
\begin{equation}
\sigma_I(\omega) = \sum_F \sigma_{I\to F}(\omega) = \frac{4}{3}\alpha \pi^2 \omega \sum_F \delta(E_F-E_I-\omega) \sum_{M=-1,0,1} \left|\langle \Psi_F^N|\hat{d}_M|\Psi_I^N \rangle \right|^2,
\label{eq:cross_section}
\end{equation}
where $\alpha$ is the fine structure constant, $\Psi_I^N$ and $\Psi_F^N$ are the initial and final molecular eigenstate with energy $E_I$ and $E_F$, respectively, $\omega$ is the photon energy, and $\delta(E_f-E_I-\omega)$ a function that determines the shape and position of the line.
The many-electron dipole-matrix element in Eq.~(\ref{eq:cross_section}) is given by 
\begin{equation}
\langle \Psi_F^N|\hat{d}_M|\Psi_I^N \rangle = 
\sum_{i,j} \langle \Psi_F^N|c^\dagger_i c_j|\Psi_I^N \rangle
\langle i| \hat{d}_M |j \rangle, \label{eq:dipole_mom}
\end{equation}
where
$\langle i | \hat{d}_M| j \rangle$  is the dipole moment operator,
$c_i$ and $c_i^\dagger$ are fermionic operators that annihilate and create an electron, respectively, 
and the sum over $i$ and $j$ runs over a complete set of one-particle basis functions.

To account for the finite width 
of the vibrational wave-packet evolving in time and following the arguments in Ref.~\citenum{marchenko_ultrafast_2018}, we employ 
\begin{equation} \delta(E_F-E_I-\omega) = \int dr_I \exp \left( \frac{ -(E_F(r_I)-E_I(r_I)-\omega)^2 }{2 \sigma^2} \right)  |\chi_I(r_I,t)|^2,  \end{equation}
where $|\chi_I(r_I),t)|^2$ is the time-dependent vibrational wavepacket density projected along the coordinate $r_I$ 
and $\sigma$ a parameter that accounts for additional broadening effects, e.g. finite lifetime and spectral bandwidth of the x-ray pulse.
In the following we choose $\sigma=\SI{1/2.355}{\eV}$.

For the calculation of the XPS, we employ the XMOLECULE toolkit\cite{hao_efficient_2015,inhester_xray_2016,inhester_chemical_2018}.
We evaluate the transition dipole matrix element in Eq.~(\ref{eq:dipole_mom}) on the basis of the one-center approximation.
To that end, the final electronic state, $\Psi_F^N$, is separated into a continuum wave function $\phi_k$ and a bound, core-ionized 
multielectron part $\Psi_F^{(N-1)}$.
The respective continuum wave function describing the photoelectron with kinetic energy $k^2/2=\omega-(E_F-E_I)$ is 
approximated with the continuum wave function 
from an atomic carbon Auger decay calculation.
We describe the bound electronic structure with a CASCI(6,4)/6-311G calculation, where
for the final electronic state $\Psi_F^{(N-1)}$, the carbon core orbital is kept singly occupied.
This is done employing an orbital basis obtained from a ROHF calculation for the neutral, excited state $^3E$ and the $^4E$ carbon-core-ionized configuration, respectively.
Because, we use different sets of orbitals for the initial and final states,
evaluation of the transition dipole moment (Eq.~(\ref{eq:dipole_mom})) requires the
calculation of overlap matrices of the involved spin configurations\cite{inhester_chemical_2018}.
We calculate the XPS for all the accessible core ionized states 
of irreducible representation $E$ that result from the 
full diagonalization of the complete active space. This results in the core ionized states $1^4E,2^4E,1^2E,2^2E,3^2E,4^2E,5^2E$, whose potential energy curves are depicted in Fig.~\ref{fig:pec} together with the potential energy curve of the initial UV-excited state $1^3E$.


For the calculation of the AES, we employ the methods described in Ref.~\citenum{inhester_chemical_2018}.
The rate for the auto-ionization of an initial core ionized state $\Psi_I^{(N-1)}$ 
with a hole in orbital $\phi_c$ to a final state $\Psi_F^{(N-1)}$ with reoccupied core hole, two holes in the valence, and an Auger electron $\phi_\epsilon$ 
is given by
\begin{equation}
\Gamma_{I \to F}= \left| \sum_{a<b}  \langle \Psi_F^{(N-1)} |c_b c_a c^{\dagger}_{\epsilon} c^\dagger_c  |\Psi_I^{(N-1)}\rangle \, 
\langle \phi_{c} \phi_{\epsilon}| \frac{1}{r_{12}} | \phi_a \phi_b \rangle \right|^2 \label{eq:Auger_CI},
\end{equation}
where
$\langle \phi_{c} \phi_{\epsilon}| 1/r_{12} | \phi_a \phi_b \rangle$ is the Coulomb matrix element
for the interaction of two valence electrons $\phi_a$ and $\phi_b$ with the continuum electron $\phi_{\epsilon}$ and the core electron $\phi_c$.
In Eq.~(\ref{eq:Auger_CI}) the sum over $a<b$ runs over all valence electron pairs $\phi_a$ and $\phi_b$.
We evaluate the two-electron integral matrix elements by employing the one-center approximation\cite{inhester_xray_2016}. In this approximation
the wave function for the Auger electron, $\phi_{\epsilon}$, is approximated with the wave function of the atomic Auger electron.
The initial state $\Psi_I^{(N-1)}$ is described the same way as the bound part of the final electronic states in the XPS calculation.
The bound electronic structure of the final states is described with a multi-reference CI calculation involving all 
two-valence-hole configurations out of the CASCI(6,4) reference space employing an orbital set obtained 
from a ROHF/6-311G calculation for the dicationic triplet ground state ($^3E$).
We assume that the molecular geometry is static during the short lifetime of the core hole ($\simeq \SIrange{5}{10}{\fs}$).
To account for the fact that we have multiple core hole states,
the resulting Auger spectrum is constructed from the product of the partial core-photoionization cross section 
and respective Auger decay rate summed over all intermediate core ionized states.
We convolute the resulting Auger spectrum with Gaussian of \SI{1}{\eV} full-width-at-half-maximum to account for 
vibrational and finite lifetime broadening effects.


\begin{acknowledgement}
The authors thank Oriol Vendrell, Markus Ilchen, Dwayne Miller, Steven Vancoillie, Valera Veryazov, and Todd Martinez for helpful discussions.
This work was supported by the AMOS program within the Chemical Sciences, Geosciences, and Biosciences Division of the Office of Basic Energy Sciences, Office of Science, U.S. Department of Energy.  Z.L. is grateful to the Volkswagen Foundation for financial support through a Peter Paul Ewald postdoctoral fellowship. T.W. thanks the German National Academy of Sciences Leopoldina for a fellowship (Grant No. LPDS2013-14).
Partial financial support from the Czech Ministry of Education, Youth and Sports, Czech Republic (grants numbers LTT17015, LM2015083) is acknowledged by N. Medvedev. 
\end{acknowledgement}

\begin{suppinfo}
Comparison of calculated XPS and AES data with available experimental data,
discussion on the impact of finite-time resolution in realistic experimental setups,
complementary analysis of the 
charge transfer upon core shell ionization employing the classical over-the-barrier model and the MCTDH simulation.
\end{suppinfo}




\providecommand{\latin}[1]{#1}
\makeatletter
\providecommand{\doi}
  {\begingroup\let\do\@makeother\dospecials
  \catcode`\{=1 \catcode`\}=2 \doi@aux}
\providecommand{\doi@aux}[1]{\endgroup\texttt{#1}}
\makeatother
\providecommand*\mcitethebibliography{\thebibliography}
\csname @ifundefined\endcsname{endmcitethebibliography}
  {\let\endmcitethebibliography\endthebibliography}{}


\end{document}


In this Supplementary Information, we present details of quantum dynamical model and a comparison of calculated data for \ce{CH3I} and methane with available experimental data.
We discuss the effect of finite time resolution on the observed signal. Further, we discuss
the classical model of charge transfer between the core ionized carbon atom and the iodine atom~\cite{erk_imaging_2014,ryufuku_oscillatory_1980}, which is adapted to our case of \ce{CH3} and I with partial charges. 

%

\section{Comparison of the calculated and experimental XPS binding energies}
Due to the lack of time-resolved experimental data for the photodissociation of \ce{CH3I}, we can only benchmark the quality of our results by comparison of the C 1s binding energy in the electronic ground state with experimental measurements. The experimental value of the binding energy is 291.43 eV \cite{bakke_table_1980}, which is in good agreement to our calculated value of 291.97 eV. 

\section{Comparison of the calculated and experimental XPS and AES spectra of methane}
\begin{figure}
\includegraphics[width=0.7\textwidth]{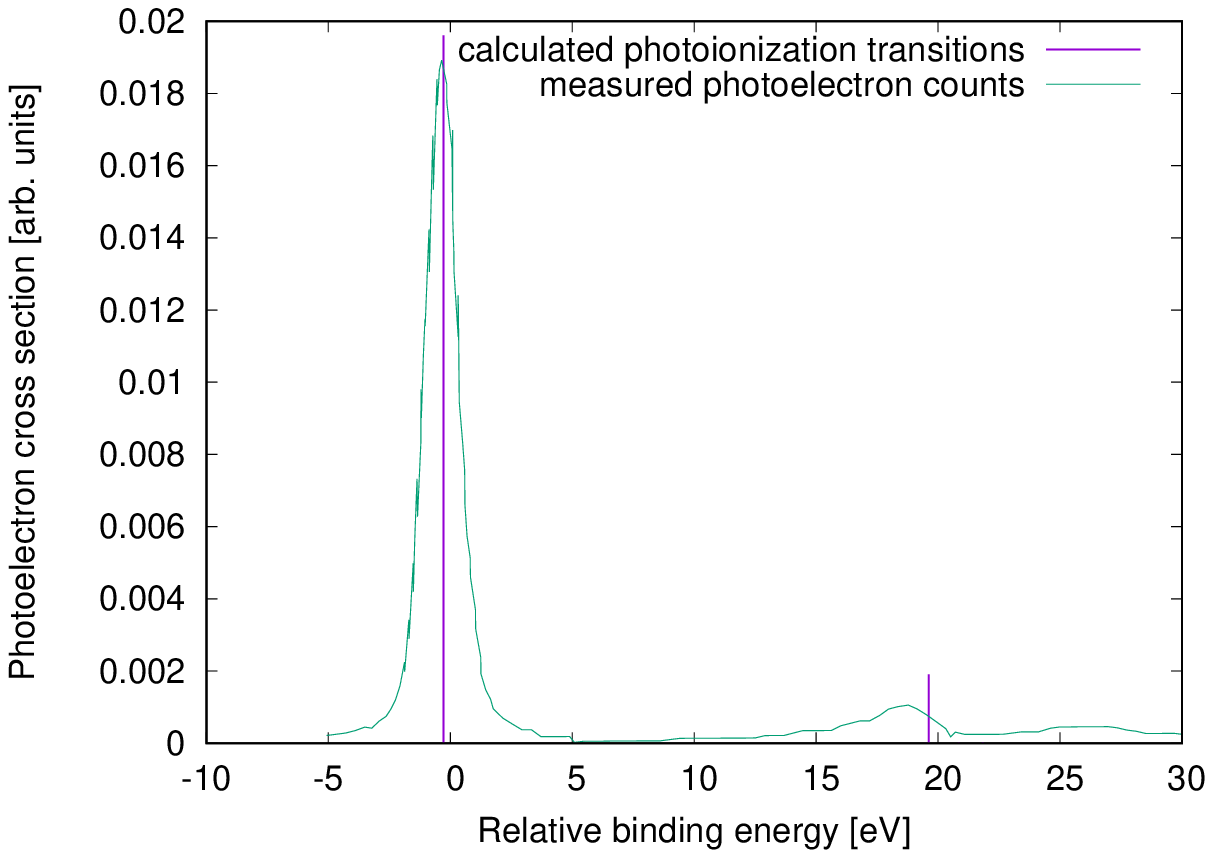}
\caption{\label{fig:methane_xps} Comparison of calculated and measured XPS. Experimental data extracted from Ref.~\citenum{creber_experimental_1980}.}
\end{figure}
To verify our calculation procedure, we compare the XPS of methane with calculations employing the same parameters (6-311G basis set, small active space comprising the lowest bound and the nearest unbound orbitals) 
in Fig.~\ref{fig:methane_xps}.
The measured data is extracted from Ref.~\citenum{creber_experimental_1980}. 
According to this reference, the main photoelectron line is at 290.7 eV, which 
is in good agreement with our calculated value of 291.2 eV.
As can be seen in Fig.~\ref{fig:methane_xps}, 
the calculated data indicates a shake-up satellite at 19.9 eV below the main line in 
good agreement with the experimental position of 18.9 eV.
The relative strength of shake-up to main line are also well reproduced in the calculation.

\begin{figure}
\includegraphics[width=0.7\textwidth]{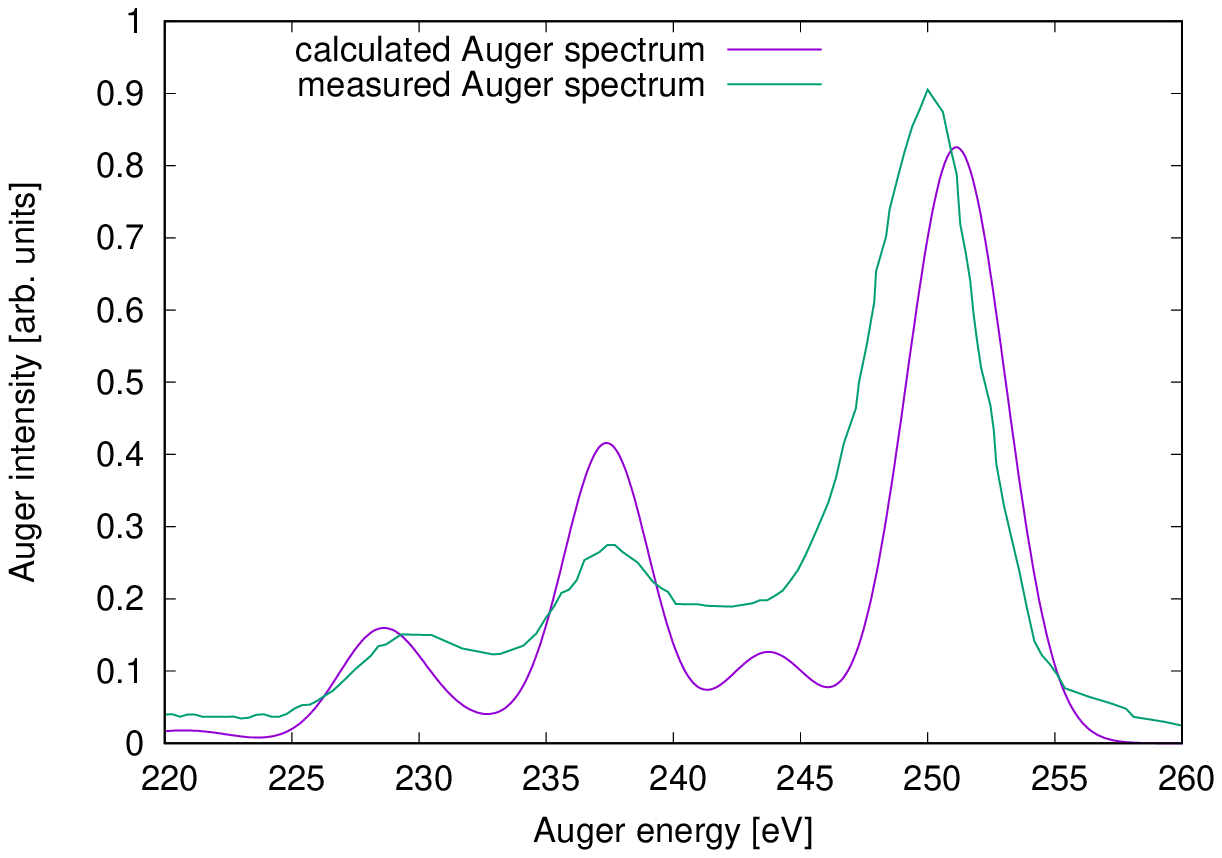}
\caption{\label{fig:methane_aes} Comparison of calculated and measured AES. Experimental data extracted from Ref.~\citenum{kivimaki_1s_1996}.}
\end{figure}
We further compare calculated AES for methane with available experimental data extracted from Ref.~\citenum{kivimaki_1s_1996}.
For a better comparison, we convoluted the spectrum with Gaussians of 4eV full width at half maximum and 
slightly shifted the calculated spectrum by 1.5 eV to lower energies. 
As can be seen in Fig.~\ref{fig:methane_aes}, the resulting spectrum reproduces all the qualitative features of the experimental Auger spectrum.

\section{XPS signature convolved with experimental time resolution}
\begin{figure}
\includegraphics[width=0.8\textwidth]{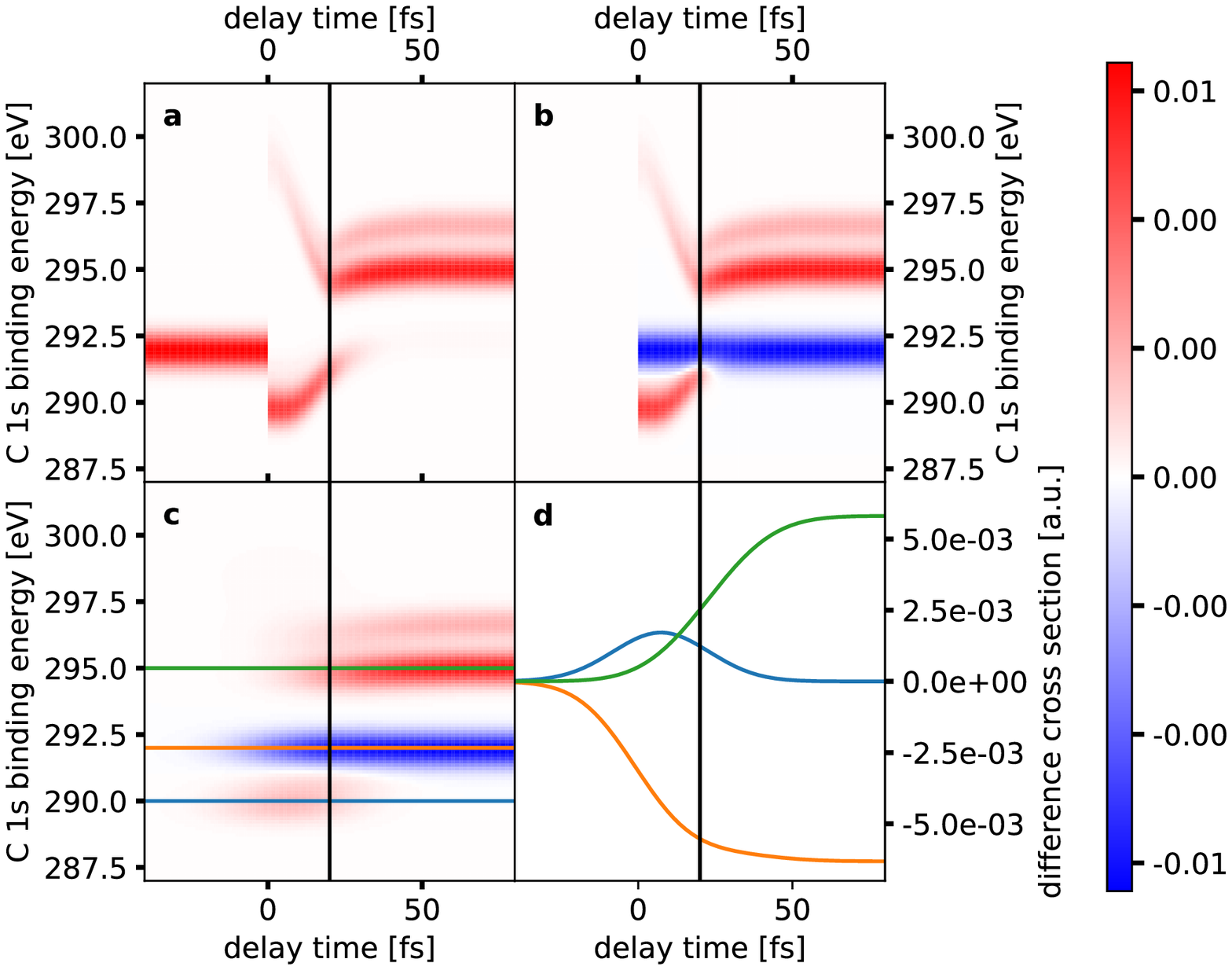}
\caption{Simulation of time-resolved XPS spectra assuming Gaussian UV and x-ray pulses with a cross-correlation of 35 fs full width at half maximum (FWHM). a) Time-resolved XPS-spectrum assuming delta functions for the time profiles of pump and probe pulse. In comparison to Fig. 2b in the main paper, the ground state XPS spectrum of \ce{CH3I} is added for negative time delays.b) Same spectrum as a), but with the ground state XPS spectrum subtracted. The ground state contribution now appears as a negative contribution at positive time delays. c) Same spectrum as b), but convoluted with 35 fs FWHM Gaussian. d) Time-dependence at different positions of the difference spectrum in c), 290 eV (blue), 292 eV (orange), and 295 eV (green).
\label{fig:Exp_XPS}
}
\end{figure}
Based on our simulations in Fig.~2b of the main paper, we simulated the signatures which would be observable in an experiment with finite time resolution. Usually, the time resolution of a optical/free-electron-laser experiments is determined by three components, the duration of the optical laser pulses, the duration of the x-ray pulses, and, either the shot-by-shot jitter of their relative timing or the precision to which this jitter can be measured and corrected for. UV laser pulses with durations $\leq$30 fs can be created without substantial effort. Many free electron lasers offer pulse durations down to $\leq$ 10 fs at appreciable intensities. The timing jitter between optical and x-ray pulses can be measured with 20 fs accuracy. In combination, this results in an achievable time resolution of ~35 fs. In Fig.~\ref{fig:Exp_XPS}a we are showing a time-dependent XPS spectrum assuming 100\% excitation and a delta function temporal profile of UV and x-ray pulses. It shows the ground state spectrum of \ce{CH3I} for negative time delays and the signatures of the bond dissociation from Fig.2b of the main paper at positive delays. Fig.~\ref{fig:Exp_XPS}b shows the result of subtracting the ground state XPS spectrum from the spectrum in Fig.~\ref{fig:Exp_XPS}a. This is a common representation of experimental data in time-resolved x-ray spectroscopy. Here the relative intensities of spectral features are independent of the ratio of photoexcited molecules vs.~molecules in the ground state. The ground state XPS spectrum now appears as a negative signature for positive time delays. In Fig.~\ref{fig:Exp_XPS}c, the difference spectrum from Fig.~\ref{fig:Exp_XPS}b is convoluted with a Gaussian of 35 fs full width at half maximum in time, which accounts for the achievable experimental time resolution discussed above. At realistic time resolutions, the shift of the strong excited state XPS feature starting at 290 eV to higher binding energies as well as the shift of the weak satellite feature starting at 299.5 eV to lower binding energies are washed out. However, the switch in XPS cross-sections from the main excited state feature at early times to the satellite feature at later times is still well observable. Fig.~\ref{fig:Exp_XPS}d shows lineouts from Fig.~\ref{fig:Exp_XPS}c. Time zero is clearly visible in the onset of the negative ground state bleach feature at 292 eV and the initially intense positive feature at 290 eV binding energy. The switch in cross-section 20 fs after UV excitation, marked by a vertical black line, is clearly visible in the decay of the 290 eV feature and the delayed onset of the positive feature at 295 eV.

\section{Classical model of charge transfer}
\begin{figure}
\begin{center}
\includegraphics[width=15cm]{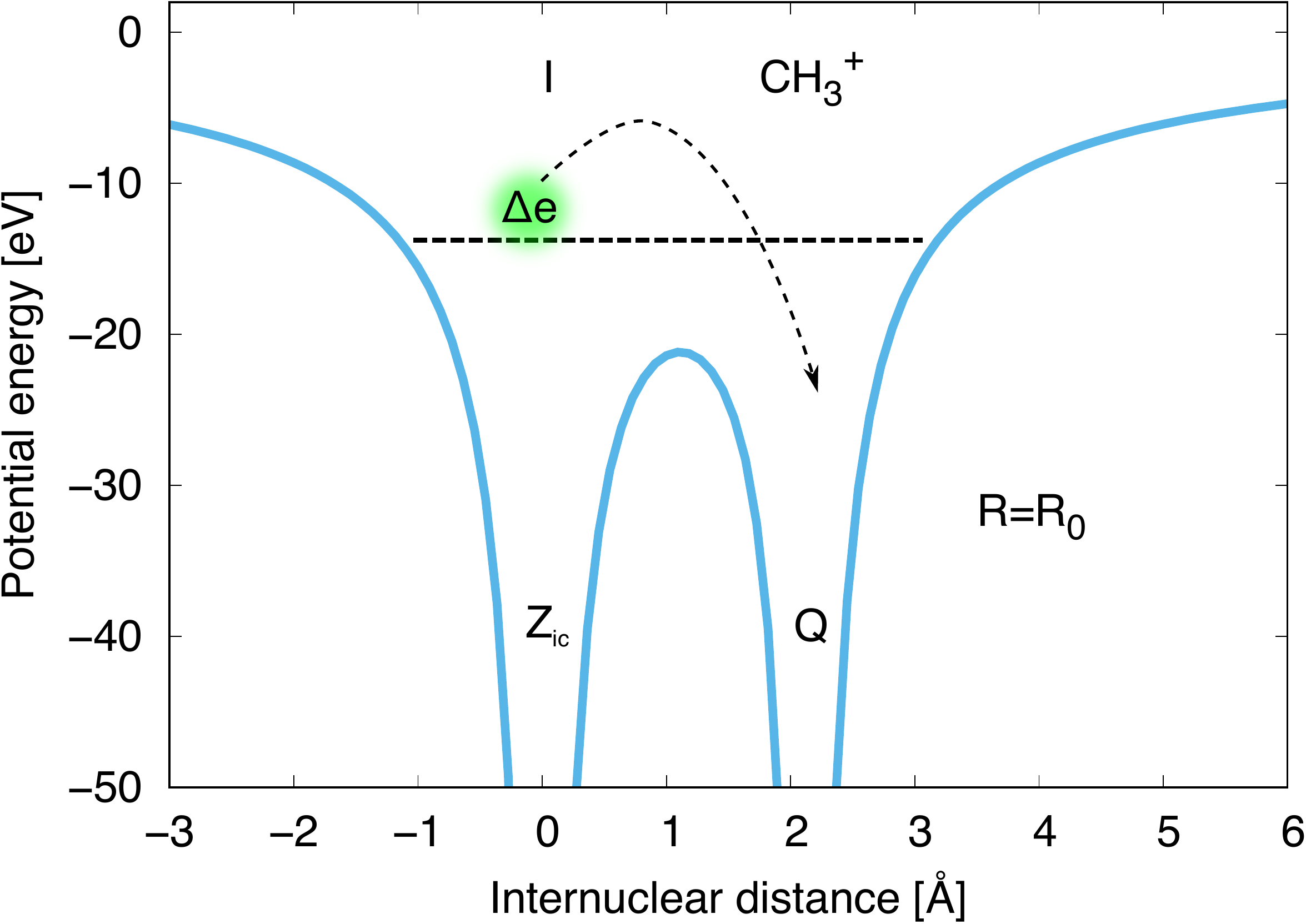}
\caption{\label{fig:FigS1} The illustration of potential and the charge (green sphere) to be transferred from iodine to the core ionized CH$_3^+$. $\zic=1$ is the charge of the ion core of I, and $Q=0.85$ is the partial charge of CH$_3^+$. The dashed line represents the energy of the lowest bound electron in I atom, which is to be transferred to CH$_3^+$, which is $E_b=-\text{IP}-|\Delta e|Q/R$, and $R=R_0=2.136$ \AA, $\De=-0.85$.
  }
\end{center}
\end{figure}

\begin{figure}
\begin{center}
\includegraphics[width=15cm]{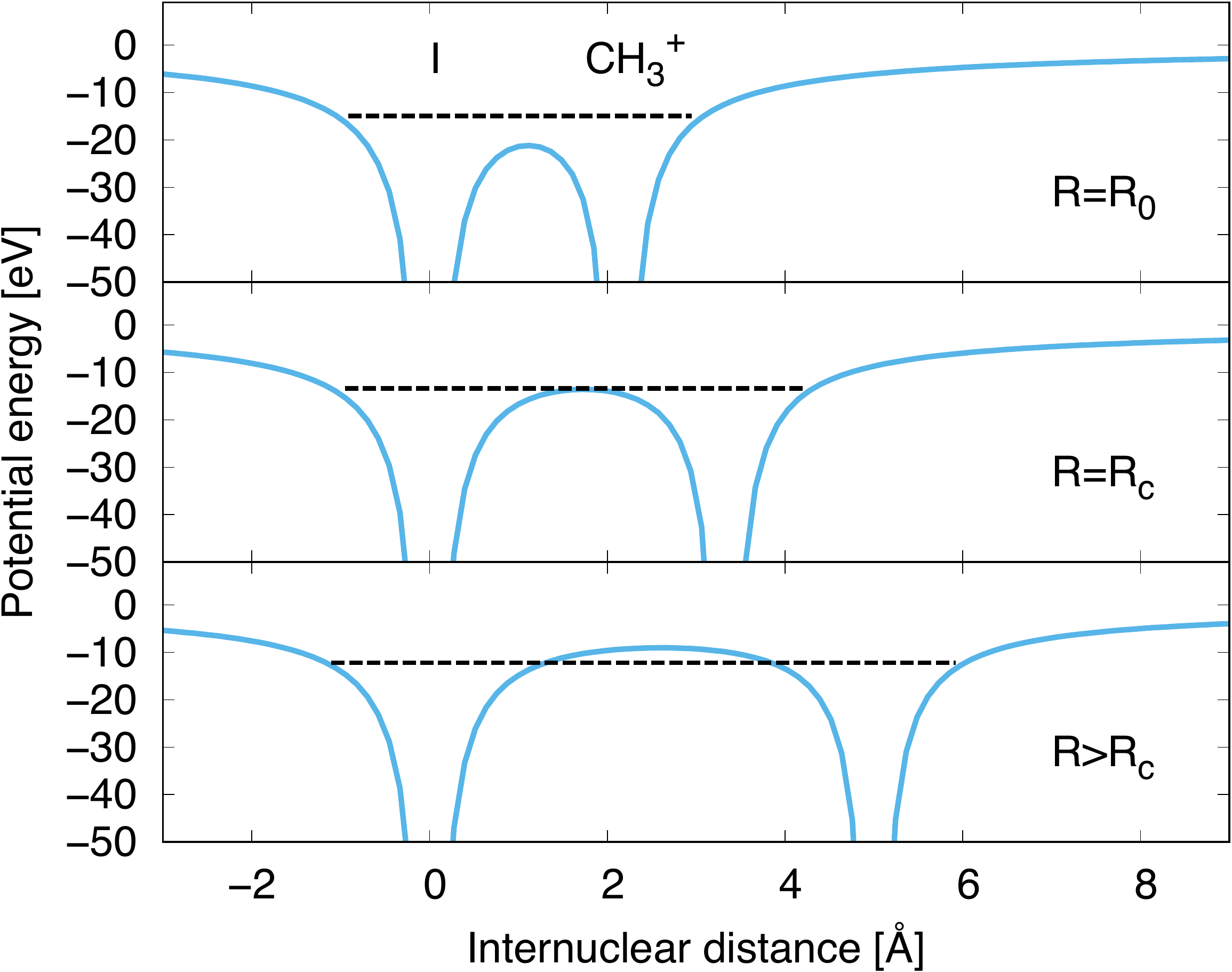}
\caption{\label{fig:FigS2} The evolution of internuclear potential experienced by the charge to be transferred from iodine to the core ionized CH$_3^+$, for internuclear distance $R$ to be the \mol equilibrium distance $R_0=2.136$ \AA, the critical distance $R_c=3.3$ \AA\ and $R=5.0$ \AA.
  }
\end{center}
\end{figure}

The CH$_3$ and I in the neutral \mol molecule have partial charges $-\delta q$ (CH$_3$) and $\delta q$ (I), where $\delta q=0.15$ is determined from the Mulliken charge analyses at Hartree-Fock level, since the cation after screening-induced charge migration is correlated with the \CH3+I$^+$ asymptote, we assume that fractional charge $\De=-0.85$ is transferred from I to \CH3, leaving behind an ion core I$^+$ of charge $\zic=1$. The partial charge of CH$_3^+$ before relaxation is $Q=1-\delta q=0.85$, and the fractional charge is transferred in the potential shown in Fig.~\ref{fig:FigS1}
\begin{eqnarray}
  \label{eq:potent}
V(x)=-\frac{|\De|\zic}{|x|} - \frac{|\De Q|}{|R-x|}
\,,
\end{eqnarray}
where $R$ is the internuclear distance between C and I located at $R$ and $0$, and $x$ is the coordinate of the moving charge modeled as point particle in the 1D system.
The potential maximal in the internuclear region $0<x<R$ is found to be at $x_0={R}/{[1+\sqrt{Q/\zic}]}$, and
\begin{eqnarray}
  \label{eq:Vm}
  V_{\text{m}}(x_0)=-|\De|\left[\frac{\zic}{R} + \frac{2\sqrt{\zic Q}}{R} + \frac{Q}{R} \right]
  \,.
\end{eqnarray}
Classically, the energy of the lowest bound electron is
\begin{eqnarray}
\label{eq:Elbe}
  E_{\text{lbe}}=-\text{IP}-\frac{|\De|Q}{R}
  \,,
\end{eqnarray}
where $\text{IP}=10.45$ eV is the 1st ionization energy of iodine atom, and the second term in Eq.~\ref{eq:Elbe} is the perturbative energy from the CH$_3^{Q+}$ ion before charge migration.
The charge migration requires naturally the condition $E_{\text{lbe}} \geq V_{\text{m}}(x_0)$ (see Fig.~\ref{fig:FigS2}), it gives the expression of the critical internuclear distance $R_c$
\begin{eqnarray}
  \label{eq:Rc}
  R\leq R_c=|\De|\left[
   \frac{\zic+2\sqrt{\zic Q}}{\text{IP}}
   \right]
  \,,
\end{eqnarray}
which is calculated to be $R_c=3.3$ \AA.


\providecommand{\latin}[1]{#1}
\makeatletter
\providecommand{\doi}
  {\begingroup\let\do\@makeother\dospecials
  \catcode`\{=1 \catcode`\}=2 \doi@aux}
\providecommand{\doi@aux}[1]{\endgroup\texttt{#1}}
\makeatother
\providecommand*\mcitethebibliography{\thebibliography}
\csname @ifundefined\endcsname{endmcitethebibliography}
  {\let\endmcitethebibliography\endthebibliography}{}